\documentstyle[aps,prb,epsfig,multicol]{revtex}

\def\bk{{\bf k}}
\def\br{{\bf r}}

\def\bR{{\bf R}}

\def\bG{{\bf G}}   

\def\bd{{\bf d}}

\def\b0{{\bf 0}}

\def\be{\begin{equation}}
\def\ee{\end{equation}}
\def\bea{\begin{eqnarray}}
\def\eea{\end{eqnarray}}
\def\beann{\begin{eqnarray*}}
\def\eeann{\end{eqnarray*}}
\def\bfig{\begin{figure}}
\def\efig{\end{figure}}
\def\btab{\begin{table}}
\def\etab{\end{table}}
\def\ba{\begin{array}}
\def\ea{\end{array}}

\begin{document}

\draft

\newcommand{\mybeginwide}{\end{multicols}\widetext
    \vspace*{-0.2truein}\noindent
    \hrulefill\hspace*{3.6truein}}
%\newcommand{\myendwide}{\hspace*{3.6truein}\noindent\hrulefill
%    \begin{multicols}{2}\narrowtext\noindent}

%\twocolumn[\hsize\textwidth\columnwidth\hsize\csname @twocolumnfalse\endcsname

%title
\title{Electronic states and optical properties of GaAs/AlAs and GaAs/vacuum
superlattices by the linear combination of bulk bands method}

%authors and addresses
\author{S. Botti and L.C. Andreani} 
\address{
Istituto Nazionale per la Fisica della Materia and 
Dipartimento di Fisica "A. Volta", \\
Universit\`a di Pavia, Via Bassi 6, I-27100 Pavia, Italy}

\date{\today}
\maketitle
\widetext

%abstract
\begin{abstract}

The linear combination of bulk bands method 
recently introduced by Wang, Franceschetti and Zunger 
[Phys.\ Rev.\ Lett.\ {\bf 78}, 2819 (1997)] 
is applied to a calculation of energy bands and optical constants 
of (GaAs)$_n$/(AlAs)$_n$ and (GaAs)$_n$/(vacuum)$_n$ 
(001) superlattices with $n$ ranging from 4 to 20. 
Empirical pseudopotentials are used for the calculation of the bulk energy bands. 
Quantum-confined induced shifts of critical point energies 
are calculated and are found to be larger for the GaAs/vacuum system.
The $E_1$ peak in the absorption spectra has a blue shift
and splits into two peaks for decreasing superlattice period; 
the $E_2$ transition instead is found to be split for large-period
GaAs/AlAs superlattices.
The band contribution to linear birefringence of GaAs/AlAs superlattices
is calculated and compared with recent experimental results 
of Sirenko et al.\ [Phys.\ Rev. B {\bf 60}, 8253 (1999)].
The frequency-dependent part reproduces the observed increase
with decreasing superlattice period, while the calculated zero-frequency
birefringence does not account for the experimental results
and points to the importance of local-field effects.

\end{abstract}

% 71.15.Hx Pseudopotential method
% 73.20.Dx Electron states in low-dimensional structures (superlattices,
%          quantum well structures and multilayers)
% 78.20.Bh Optical properties: Theory, models, and numerical simulation
% 78.20.Fm Optical properties:  Birefringence
% 78.66.Fd Optical properties of specific thin films, surfaces, and low-
%          dimensional structures... : III-V semiconductors

%] \newpage%%%%%%%%%%%%%%%%%%%%
% \leftskip 54.8pt
% \rightskip 54.8pt
%%%%%%%%%%%%%%%%%%%%

%%%%%%%%%%%%%%%%%%%%
\begin{multicols}{2}
\narrowtext
%%%%%%%%%%%%%%%%%%%%

%introduzione
\section{introduction}

Quantum confinement in semiconductor heterostructures modifies
the energy and dimensionality of electronic levels and leads
to blue shifts of the optical gaps. Most experimental
investigations have focused on the energy region of the
fundamental gap, which is easily accessible by photoluminescence
and photoluminescence excitation spectroscopies and yields
a variety of interesting physical phenomena related to bound
excitonic states; relatively few studies of confinement effects
on high-energy transitions have been presented. Concerning theory,
confined electronic levels close to band edges and the resulting
optical properties can be calculated rather simply and accurately
by the envelope-function method. The theoretical problem of 
determining optical spectra of semiconductor heterostructures in the
whole visible region is much more complex and beyond the reach
of effective-mass methods, as it requires a description of the effects 
of confinement on electronic states in the whole Brillouin zone.

Interband absorption spectra of tetrahedral semiconductors 
are dominated by two prominent features, 
denoted $E_1$ and $E_2$.\cite{chelikovsky-cohen,yu-cardona}
The $E_1$ peak (and its spin-orbit counterpart $E_1+\Delta_1$)
originates from band-to-band transitions along the $\Gamma-L$ direction, 
where valence and conduction bands are nearly parallel:
this results in a $M_1$-type critical point, i.e., a saddle point
in the joint density of states, which also gives a strong
excitonic character to the transition.
The $E_2$ peak, instead, has contributions from different parts
of the Brillouin zone, but mainly from a region centered around
the special point $({3\over4},{1\over4},{1\over4})$
(in units of $2\pi/a$, where $a$ is the lattice constant).
The $E_2$ peak has essentially no excitonic character.
Blue shifts and splittings of the $E_1$ and $E_2$ transitions were
measured in GaAs/AlAs superlattices \cite{garriga87,alouani88}.
More recently, a quantum confinement induced shift of $E_1$ and $E_2$ 
was measured in Ge nanoparticles embedded 
in a glassy matrix \cite{tognini_prb,apl_paper}.

A main purpose of this work is to calculate the behavior of $E_1$ and $E_2$
transitions upon confinement in GaAs/AlAs superlattices and in 
free-standing GaAs layers, which are simulated by GaAs/vacuum superlattices.
Besides the obvious interest of the GaAs/AlAs system, the motivation
for choosing GaAs/vacuum superlattices is to study confinement effects
on $E_1$ and $E_2$ in a system where the electronic states are truly
confined in the GaAs layers even at high energies. 
In GaAs/AlAs superlattices, on the other hand,
the band structures of the two constituents far away from the fundamental
band edges are rather similar and strong banding effects occur 
in short-period structures, i.e., the electronic states become 
delocalized along the superlattice. A comparison between the two 
systems should therefore elucidate the respective roles of quantum
confinement and superlattice band formation in determining the
optical properties. Free-standing GaAs films can be 
produced by chemical etching \cite{gerard};
GaAs/vacuum superlattices can also be a model for multilayers formed
by GaAs and a wide-gap oxide, like Al$_2$O$_3$ or oxidized AlAs (AlOx).

The electronic structure~of~short-period (GaAs)$_n$/ (AlAs)$_n$ superlattices
can be calculated from first principles by density-functional theory
and norm-conserving pseudopotentials \cite{baldereschi88}.
The computing time grows rapidly with the number $n$ of
monolayers.\cite{note_period}
Moreover, density-functional calculations
of the optical properties suffer from the gap problem \cite{gap_problem}
and require the inclusion of quasiparticle corrections \cite{quasiparticle}
and excitonic effects,\cite{excitonic}
thus increasing the computational cost.
Recently a new pseudopotential approach has been developed \cite{wang97},
in which the electronic states of a heterostructure are represented
as linear combinations of bulk bands (LCBB). This method can be applied
to heterostructures with a large number of atoms per primitive cell
with a reasonably low computational effort; it is particularly suited to
study how the optical spectra of the bulk are modified upon confinement,
since the electronic states of the bulk are the starting point 
of the technique. The LCBB method is adopted in this work. 
The results presented here can be the starting point for tackling 
more complex problems (e.g. the inclusion of local-field and excitonic 
effects, or the study of confinement in large quantum dots) as well as 
for comparison with ab-initio calculations when they will become available.

While GaAs and AlAs are cubic materials and have an isotropic
dielectric constant, (001)-grown superlattices are intrinsically
uniaxial and the component $\epsilon_{\parallel}$ of the dielectric tensor
parallel to the growth direction differs from the perpendicular
component $\epsilon_{\perp}$. The linear birefringence
$\Delta n=(\epsilon_{\perp})^{1/2}-(\epsilon_{\parallel})^{1/2}$
has been measured in GaAs/AlAs superlattices \cite{fainstein94,sirenko99}:
$\Delta n$ is of the order of a few percent and has a nontrivial
dependence on the superlattice period.
Birefringence is much larger in GaAs/AlOx multilayers,
where it has been employed to achieve phase matching for
second-harmonic generation \cite{fiore98}. A goal of the
present work is to calculate linear birefringence \cite{note_birefringence}
and its dependence on energy and superlattice period.
The calculations presented here yield only the contribution
to birefringence arising from quantum-confinement-induced modifications
of the electronic states; they do not account for the intrinsic
dielectric anisotropy of a multilayer arising from different
boundary conditions for an electric field parallel or perpendicular
to the layers \cite{agranovich85,cardona93}. This second contribution
to birefringence is in fact equivalent to the inclusion of local-field
effects in the dielectric response \cite{adler61,wiser62,hanke78,nazarov94}
and is left for future work.

The rest of this paper is organized as follows. In Sec.\ II 
we briefly describe a few aspects of the adopted method: the LCBB technique 
combined with empirical pseudopotentials,\cite{wang97,zunger99}
as well as space symmetry and the tetragonal Brillouin zone of 
the superlattice.
In Sec.\ III we present the results for GaAs/AlAs and GaAs/vacuum 
superlattices. These are divided into three parts: electronic levels
(with the added complication of surface states in the GaAs/vacuum system),
optical properties via the real and imaginary parts of the dielectric
function, and birefringence.
Section IV contains a discussion of the present findings also
in view of extensions of this work.

%%%%%%%%%%%%%%%%%%%%%%%%%%%%%%%%%%%%%%%%%%%%%%%%%%%%%%%%%%%%%%%%%%%%%%%%%%%%%%

\section{method of calculation}

We consider (GaAs)$_n/$(AlAs)$_n$ and (GaAs)$_n/$(vacuum)$_n$ superlattices
grown along a (001) crystallographic surface
and with period $d=na$ with $n$ ranging form 4 to 20.
The ideal structure for a lattice-matched system with abrupt interfaces 
is a simple tetragonal Bravais lattice, 
with a supercell defined by the basis vectors 
$ \left( 1,1,0 \right) a/2 $, $ \left( -1,1,0 \right) a/2 $, 
$ \left( 0,0, 1 \right) na $, where $a$ is the bulk lattice constant. 
The reciprocal lattice is also simple tetragonal, with basis vectors 
$ \left( 1,1,0 \right) (2\pi)/a $, $\left( -1,1,0 \right) (2\pi)/a $
and $ \left( 0,0,1 \right) (2\pi)/(na)$. The first Brillouin zone is 
shown in Fig. 1. Superlattice high symmetry points are distinguished 
from their bulk counterparts by putting a bar over the symbol.
An additional symmetry point $\bar L$ is defined as follows:
$\bar L=\bar X$ if $n$ is even, $\bar L=\bar R$ if $n$ is odd. 
The most important zinc-blende $\bk$ points are folded onto 
superlattice points as follows:
\bea
 & &  \Gamma , \left\{ \frac{j}{n} X^z \right\}_{j=-n+1,n} 
 \, \; \; \; \; \longrightarrow  \bar \Gamma \,, 
 \nonumber \\
 & &   X^y , \left\{ \frac{j}{n} X^z \right\}_{j=-n+1,n}
 \;  \, \longrightarrow  \bar M \,,
 \nonumber \\
 & &  L_{111} , \left\{ \frac{j}{n} X^z \right\}_{j=-n+1,n}
  \longrightarrow  \bar L \,.
\eea
In the case of a common anion structure like (GaAs)$_n/$(AlAs)$_n$
the point group is $D_{2d}$, otherwise it is $C_{2v}$: the latter is the
case of (GaAs)$_n/$(vacuum)$_n$ superlattices.\cite{sham-lu}

In order to study large scale systems, we adopt a 
LCBB method \cite{wang97,zunger99}:  
superlattice electronic wave functions are expressed
as linear combinations over band indices $n_b$ and wave vectors $\bk$
of full-zone Bloch eigenstates of the constituent bulk materials.
Because of the requirement of periodicity, 
the superlattice potential mixes only bulk states 
labelled by $\bk$ vectors which differ by a superlattice reciprocal 
lattice vector $\bG_{SL}$: 
the number of coupled states is hence always equal to $2n$, 
because exactly $2n$ $\bG_{SL}$ are contained in the fcc Brillouin zone.  
The maximum dimension of the basis set is then given by $2n$ 
multiplied by the number of selected bulk bands indices.
In the case of a GaAs/vacuum superlattice we decide to 
include the 4 valence bands and the 4 lowest conduction
bands in the basis set.
In the case of a GaAs/AlAs superlattice the roughest selection is to take both
GaAs and AlAs bulk states at each mixed $\bk$ and $n_b$, 
orthonormalizing at the
end the basis set obtained. 
But, as GaAs and AlAs band structures are very similar 
except for the lowest conduction band
(see Fig.~\ref{bulkbands} and details of calculation below),
we have verified that it is enough to  include only GaAs states for $n_b$
from 1 to 8 together with the 5th band of AlAs (i.e., the lowest
conduction band). The resulting set must be orthonormalized.
It can easily be seen that the final dimension of the basis is 
always small ($40 \times 9$ for the largest supercell). 
A convergence test is presented at the beginning of Sec.~III.

Unlike tight-binding or standard plane wave expansions, 
the LCBB method allows the intuitive pre-selection of physically
important states to be included in the basis set. 
By contrast with the ${\bf k} \cdot {\bf p}$ approach,
off-$\Gamma$ states $u_{n,{\bf k} \neq 0}$ are directly considered,
permitting a correct treatment of $\Gamma$-$X$ coupling
and of all confinement-induced couplings within the Brillouin zone,
without the need for a large basis of ${\bf k} = 0$ bulk states.
The formalism applied here is strain-free, believing that the small 
lattice mismatch ($\simeq 0.15 \%$) justifies a description of the 
geometrical structure as unrelaxed, with a lattice constant of 5.655 \AA\
averaged
over the constituent materials. 
All trends in the superlattice states obtained 
by LCBB method were shown to be 
reproduced \cite{wang97}, 
with a surprising accuracy (10-20 meV) and a small computational effort,
down to the monolayer superlattice and up to large periods.

A local empirical pseudopotential approach
is chosen to perform the band structure calculations, both for the 
heterostructures and the constituent compounds.
Since the adopted pseudopotentials are designed 
for a kinetic-energy cutoff of 5 Ry, \cite{mader}
bulk eigenfunctions are expanded on a basis set made of 
about 60 plane-waves at each $\bk$ point.
In Fig.~\ref{bulkbands} we show the band structures of GaAs and AlAs
calculated with these pseudopotentials.

The potential term in the one-particle Hamiltonian is built as a 
superposition of screened, spherical atomic pseudopotentials
$v_{\alpha}$:
\begin{equation}
 H=-\frac{\hbar^2\nabla^2}{2m} + 
  \sum_{\alpha} \sum_{{\bf R} \in DL} 
  \, v_{\alpha} \left( {\bf r-R-d}_{\alpha} \right)
  \, W_{\alpha} \left( {\bf R} \right) \,,
\end{equation}
where $\bR $ is a fcc direct lattice (DL) vector and $\bd_\alpha $ the 
displacement of the atom of type $\alpha $ in the bulk primitive cell. 
The weight function
$  W_{\alpha} \left( {\bf R} \right) $ 
selects the atom basis which lies on each lattice site,
describing the geometrical details and the symmetry of the structure:
in the vacuum layers its value is zero.
The Hamiltonian matrix elements on the bulk basis set depend on the 
Fourier transform of the pseudopotentials 
$v_{\alpha} \left( \br \right)$; details on the LCBB method 
can be found in Refs.\ \onlinecite{wang97,zunger99}.
We use the continuous-space functions $v_{\alpha} \left( \bk \right)$ 
proposed by M\"ader and Zunger \cite{mader}.
In  Ref.~\onlinecite{mader}
the empirical parameters are adjusted in order to fit both the 
measured electronic properties of bulk GaAs and AlAs and some local 
density approximation (LDA) results for superlattices. 
It has been verified that the wave functions of bulk and superlattices 
systems calculated with these pseudopotentials are close to those obtained 
in rigorous first principles LDA calculations \cite{mader}.   
The As potential depends on the type of nearest neighbors (Ga or Al),
thereby describing the local-environment dependence of the atomic potential,
by considering the local electronic charge. 
These pseudopotentials are adjusted to reproduce
the experimental  GaAs/AlAs valence-band offset (0.50 eV).
Bulk and superlattice energy levels are provided in the same
absolute energy scale, thus superlattice and bulk eigenvalues can be
compared directly.  
The inclusion of non-local terms is needed to achieve a better 
description of high energy states, nevertheless we have verified that
almost only transitions to the levels originated by the two lowest conduction 
bands 
are responsible for the structures in the optical spectra below 6 eV, 
therefore non-local terms can be neglected for our purposes.
At this stage we have also decided to neglect spin-orbit interaction, 
even if in Ref. \onlinecite{mader} it is suggested how to include it.
To preserve a correct description of interfaces in GaAs/AlAs superlattices,
an As atom bound to two Al and two Ga atoms has been attributed 
a symmetrized pseudopotential which is the average
of the As pseudopotential functions in GaAs and AlAs environments.
In GaAs/vacuum superlattices, the dangling bonds at the interfaces are
responsible for the appearance of surface states,
lying in the forbidden energy gap.
The sizeable metallic contribution to low energy transitions is an
artifact of the calculation (in a real system dangling bonds
would be saturated by the microscopic nature of the interface)
and must be minimized by excluding surface states from initial and
final states in optical transitions.

Starting from the one-electron band structure, the complex dielectric function
$\epsilon(\omega)=\epsilon_1+i\epsilon_2$ 
is evaluated in a straightforward way by means of 
semiclassical theory of interband transitions.\cite{yu-cardona,baspas} 
Tetragonal 
Brillouin zone integrations are performed using Fourier quadrature 
with 1056 Chadi and Cohen special points
in the irreducible wedge.\cite{ccsp,froyen}
An empirical gaussian broadening of 0.1 eV has been introduced to obtain 
smoothed curves. 
Bulk spectra have been preliminarly calculated integrating over both 
a fcc Brillouin zone 
and the corresponding folded tetragonal Brillouin zone,
verifying the exact equivalence of the two results.

In Fig.\ \ref{e2bulk} we show the calculated real and imaginary parts of
dielectric functions for bulk GaAs and AlAs crystals. Agreement with the
experimental positions of the peaks\cite{chelikovsky-cohen,yu-cardona} is
within a few tenths of an eV: we underline that the empirical approach is
free of the band gap problem.  The height of the peaks, especially $E_1$,
cannot be correctly estimated without the inclusion of the excitonic
contributions and the local field effects.

%%%%%%%%%%%%%%%%%%%%%%%%%%%%%%%%%%%%%%%%%%%%%%%%%%%%%%%%%%%%%%%%%%%%%%%%%%%%%%

\section{results}

%%%%%%%%%%%%%%%%%%%%%%%%%%%%%%%%%%%%%%%%%%%%%%%%%%%%%%%%%%%%%%%%%%%%%%%%%%%%%%
\subsection[A]{Electronic levels}

After calculating the single-particle eigenstates of both bulk systems in
the whole Brillouin zone, in order to extract the bulk states basis on
which to construct the superlattice Hamiltonian, we applied the method 
described in section II to obtain the single-particle band structures of
(001) (GaAs)$_n$/(AlAs)$_n$ and (GaAs)$_n$/(vacuum)$_n$ superlattices. 
The period $d=na$ has been varied from $4a$ to $20a$. 

When a sufficiently large number of bulk states is used 
as a basis set for the LCBB method, the results must converge to those
obtained with a direct diagonalization of the Hamiltonian
for the corresponding number of plane waves.
A comparison of LCBB results with the conventional supercell approach
was presented in Ref.~\onlinecite{wang97}.
Here we perform a convergence test, which consists in calculating
the energy levels with four different bulk basis sets of increasing size.
Selected results are shown in Tab.~\ref{tabella} for
(GaAs)$_{10}$/(AlAs)$_{10}$ and (GaAs)$_6$/(AlAs)$_6$ superlattices.
As far as valence states are concerned, dependence of the energy 
levels on the basis set is below 10$^{-3}$~eV; for the lowest 
conduction band levels the dependence on the basis set
is generally below 0.05~eV, and falls below 10$^{-2}$~eV
when the 5$^{\mathrm{th}}$ band of AlAs is included in the basis.
The results of Tab.\ref{tabella} justify the use of basis 3),
namely $n_b=1$ to 8 for GaAs and $n_b=5$ for AlAs, as stated in Sec.~II.

In Figs.~\ref{SLbandsgaal} and \ref{SLbandsvacu} we show the 
superlattice energy bands for $n=10$: the electron energy
levels are plotted  along the highest symmetry lines in the tetragonal
Brillouin zone (see Fig.\ \ref{tetrbz}).
Since $2n$ $\bk$ points in the fcc Brillouin zone are always folded 
onto the same $\bk$ point in the smaller tetragonal Brillouin zone,
the number of occupied 
superlattice bands is $2n$ times the number of bulk bands for GaAs/AlAs,
$n$ times the number of bulk bands for GaAs/vacuum superlattices.
The dispersion along the $\bar \Gamma - \bar M$ direction is similar
to the dispersion along the $\Gamma-X$ direction in the bulk,
while the other two directions $\bar\Gamma-\bar R$ and $\bar\Gamma-\bar X$
have no counterpart in the band structures of Fig.~\ref{bulkbands}.
In GaAs/AlAs, the dispersion along the growth direction 
$\bar \Gamma$-$\bar Z$ is much smaller than in the other directions, 
as expected for superlattice minibands;
in GaAs/vacuum the bands along $\bar \Gamma$-$\bar Z$ are flat 
as tunneling through the vacuum has a negligible effect.

The main differences in the superlattice band structures compared to the bulk 
can be interpreted in terms of zone folding and quantum confinement effects; 
it is also interesting to compare the band structures of the two superlattices.
Superlattice gaps are larger than bulk gaps: in particular GaAs/vacuum
gaps are larger than GaAs/AlAs ones, as a result of a stronger confinement; 
moreover superlattice band gap widths increase as the superlattice 
period decreases.
The lowering in the crystal symmetry is responsible for the removal of level
degeneracies: as an example in the GaAs/AlAs $D_{2d}$ superlattice 
the threefold degenerate valence states at $\Gamma$ (spin-orbit is
neglected) are split in a twofold-degenerate and a non-degenerate state,
while in the GaAs/vacuum $C_{2v}$ superlattices the degeneracy is completely
removed. 

In GaAs/vacuum bands we clearly see the appearance of states lying in the
forbidden energy gaps. The lowest one lies in the gap from -10 to -6 eV,
while two other ones lie in the optical gap from 0 to about 2 eV.
A fourth state can be recognized at -5 eV around the $\bar M$ point,
while in other regions of the Brillouin zone it resonates with the energy
bands. Indeed, four surface states or resonances are expected from the
presence of two dangling bonds at the two interfaces of each GaAs layers.
We can identify the surface states by studying the behavior of the
probability $\left| \psi \right|^2$ to find an electron along the growth
direction $z$, averaged over the in-plane $x,y$ coordinates.
Taking as an example the conduction miniband  states at $\Gamma$, where
the potential profile is characterized by 0.5 eV deep wells in GaAs layers, 
we observe (see Fig \ref{surface}) that an electron in a surface state
($j=39$ in the exemplified case) has a high probability to be localized on
the surface and a decaying probability to enter the GaAs layer; on the
other hand an electron in a bulk state ($j=36$ in the figure) has an
oscillating probability to be found in the GaAs layers.
Both states are obviously evanescent in vacuum.
From the position of the uppermost occupied band ($j=40$) it follows that 
GaAs/vacuum superlattices have a metallic behavior: this is an 
artifact of the calculations,
as our aim is to simulate insulating
multilayers made of GaAs and a wide-gap oxide.
We get rid of the problem by excluding surface states as initial or
final states in interband transitions. This simulates the formation of
interface states or defects in a GaAs/oxide superlattice, which would
saturate the dangling bonds.  The surface resonance cannot be easily
eliminated, but it produces small effects on the spectra, since it lies
deep in the valence band.

%%%%%%%%%%%%%%%%%%%%%%%%%%%%%%%%%%%%%%%%%%%%%%%%%%%%%%%%%%%%%%%%%%%%%%%%%%%%%%
\subsection[B]{Optical spectra}

The calculated imaginary part of the dielectric function
$\epsilon_2 \left( \omega \right)$ for both GaAs/AlAs and 
GaAs/vacuum systems are shown in Fig. \ref{eps2SL}
for different superlattice periods $n$. Here we average over the three 
orthogonal polarization directions to obtain a scalar dielectric function.
The electronic bands of GaAs and AlAs are similar and therefore 
the two optical spectra present the same features, 
namely $E_1$ and $E_2$ peaks (see Fig.\ \ref{e2bulk}).
Superlattice spectra show also these structures, which can be  
compared directly to bulk spectra: in fact, even if in a superlattice 
selection rules allow more transitions as a consequence of
zone-folding, the tetragonal Brillouin zone is smaller and, at last, 
all transitions which contribute to superlattice peaks have their 
equivalent counterparts in the bulk Brillouin zone.
As a general remark, the $E_1$ transition is found to blue shift
and to split into two peaks for decreasing superlattice period;
the confinement-induced shift is larger for the GaAs/vacuum system.
On the other hand the $E_2$ transition is split for large-period
GaAs/AlAs superlattices, where the electronic states are confined
in the two bulk layers leading to a superposition of the two bulk spectra;
the two peaks merge into a single one for small period. A single $E_2$ 
peak with a small blue shift is found for GaAs/vacuum superlattices.

In Fig. \ref{picchi} the peak energies are plotted as a function of the 
superlattice period $n$. First we comment on the behavior of $E_1$: in zinc-blende 
crystals it comes from transitions along the $\Lambda $ line, 
in a region where bulk bands are almost parallel. 
When the system is confined in the [001] direction, it is not intuitive to
describe the consequences of folding of $\langle111\rangle$ directions.
The calculated spectra show that along the folded $\Lambda$ line 
transitions subdivide in two main groups and lead to a splitting
of the $E_1$ peak in the absorption curves. 
The two peaks have different oscillator strengths and, 
except for an intermediate period length, the lowest energy one 
becomes much stronger and covers the other one.
Both peaks undergo confinement effects and are moved towards higher
energies in comparison with their bulk position:
confinement and the consequent shifts are stronger at smaller well widths.
A splitting of the $E_1$ transition with a blue shift of both peaks
was indeed observed in GaAs/AlAs superlattices \cite{garriga87}.
In the present calculation this is attributed to a splitting
of the bulk valence band at the $L$ point and along the $\Lambda$ line,
as indicated by the band energies.
The results of Figs. \ref{eps2SL} and \ref{picchi} show also
that $E_1$ peak displacements are more relevant in GaAs/vacuum superlattices, 
where quantum confinement effects are stronger due to the vacuum barrier.

The behavior of the $E_2$ peak is substantially different: 
its main contribution comes from transitions in a region 
close to the special point $\bk= {2\pi\over a} 
\left({3\over4},{1\over4},{1\over4} \right)$.\cite{e0prime}
At this point the alignment of both valence and conduction
GaAs and AlAs bands is almost flat 
and the electronic wave functions are completely delocalized 
all over the heterostructure. 
This explains why $E_2$ peak in GaAs/AlAs superlattices is at an intermediate energy 
between bulk GaAs and AlAs $E_2$ peak positions and does not shift
when the superlattice period $n$ decreases. 
Our calculated peak positions are in good agreement with experimental
data,\cite{mendez81,garriga87,alouani88} in particular a splitting of
$E_2$ is reported in Ref.~\onlinecite{alouani88}.
In GaAs/vacuum superlattices the situation changes: electrons near the 
special point are confined in GaAs layers 
and the superlattice $E_2$ peak has a weak blue shift at small superlattice periods $n$,
going back to the bulk GaAs $E_2$ energy when $n$ grows. A single peak 
obviously arises in this case since there is no AlAs contribution.

At last, we present in Fig.~\ref{e1} some curves for the real
part of dielectric function $\epsilon_1$  for
GaAs/AlAs  heterostructures. We observe that the average, or Penn gap
(defined as the energy at which $\epsilon_1$ goes through zero) does not
depend on the superlattice period. This proves that the center of gravity of valence and 
conduction bands is preserved, as suggested in 
Ref.~\onlinecite{sirenko99}: this follows from compensating effects of a
blue shift at the bottom of the band (positive curvature) and a red shift 
at the top of the band (negative curvature). 

%%%%%%%%%%%%%%%%%%%%%%%%%%%%%%%%%%%%%%%%%%%%%%%%%%%%%%%%%%%%%%%%%%%%%%%%%%%%%%
\subsection[C]{Birefringence and absorption anisotropy}

In (GaAs)$_{n}/$(AlAs)$_{n}$ (001) superlattices 
the $T_d$ point group of the zinc-blende structure
is replaced by the $D_{2d}$ symmetry group.
Cubic crystals present isotropic optical properties, while in a
superlattice the reduction in symmetry leads to optical anisotropy
in the real part of the dielectric constant (birefringence) 
and in the imaginary part (absorption anisotropy, or dicroism). 
The system is uniaxial, with the optical axis directed along 
the growth direction $z$, thus the dielectric tensor has the form:
\be
 \epsilon_{ij} \left( \omega \right)= 
 \epsilon_{ii} \left( \omega \right) \, \delta_{ij} \,,
\ee 
where $\epsilon_{xx}=\epsilon_{yy}=\epsilon_{\perp}$ and 
$\epsilon_{zz}=\epsilon_{\parallel}$.\cite{inplane}
The macroscopic dielectric tensor including local-field effects
should be found by the procedure of inverting a
matrix $\epsilon({\bf G},{\bf G}')$ with off-diagonal elements
depending on the reciprocal lattice 
vectors ${\bf G}$.\cite{adler61,wiser62,hanke78,nazarov94}
In bulk semiconductors the off-diagonal terms have a small effect
on optical properties, in superlattices however they are expected
to give a sizeable contribution and to depend on the 
superlattice period.\cite{photonic}
Nevertheless, at this stage we do not include  
local field effects in our calculation:
we directly take into account only the effects of electronic
confinement and band folding on optical transitions. Comparison
with birefringence data reported in Ref.~\onlinecite{sirenko99} 
should allow to  determine if one of the two contributions to anisotropy
is dominant.

In Fig. \ref{anis} we present the frequency dispersion of
$\epsilon_{\parallel}$ and $\epsilon_{\perp}$, both for real and imaginary
parts: the birefringence is dispersionless up to energies close to the
direct gap,  while at higher energies it presents resonant contributions.
We see as expected that folding and confinement can have a greater 
influence on the resonant part of the birefringence:
indeed, transitions from valence subbands couple differently with
$xy$ or $z$ polarized electric fields.
Note that the interband absorption edge is higher in energy for
$z$-polarized light: this is in agreement with well known
quantum well and superlattice physics, in which the lowest transition
is a heavy hole one and is forbidden for light polarized 
along $z$.\cite{qw}
Once again the effect is greater when confinement is stronger (small
superlattice period). There is a dispersionless contribution to birefringence
at low frequencies of the order of $10^{-3}-10^{-2}$ that cannot be
distinguished in Fig.~\ref{anis}.
As proposed in Ref.\onlinecite{sirenko99}, we decouple this
low energy background birefringence, describing 
$\Delta n \left( \omega \right)
=(\epsilon_{\perp})^{1/2}-(\epsilon_{\parallel})^{1/2}$ 
in terms of a fitting function 
\be
 \Delta n \left( \omega \right) = \Delta n_{bg} - \Delta n_{gap} 
 \ln \left( 1 - \left( \frac{\omega}{\omega_g} \right)^2 \right)  \,,
\label{fit}
\ee
where $\Delta n_{bg}$ is the background contribution we want to isolate, 
the second term refers to the resonant contribution and $\omega_g$ is the
gap frequency. In Ref.~\onlinecite{sirenko99} the three parameters 
$\Delta n_{bg}$, $\Delta n_{gap}$ and $\omega_g$ are extracted by fitting
with expression (\ref{fit}) the experimental data. We also fit our
calculated  $\Delta n \left( \omega \right)$ curves 
by means of expression (\ref{fit}).
In Fig.~\ref{fitpara} we display the fit parameters as a function of the 
well width: the graphs can be easily compared with the analogous
experimental curves presented in Ref.~\onlinecite{sirenko99}.
The gap frequencies we extract by the fit agree both with directly calculated
gaps (barely visible in Fig.\ref{eps2SL}) and with the measured
ones.\cite{sirenko99} 
$\Delta n_{gap}$ shows an increase of the resonance for small periods:
the theoretical curve reproduces the trend of the experimental curve,
although the calculated values are smaller.
The sudden decrease of the measured background birefringence 
$\Delta n_{bg}$ below 40~\AA, on the other hand, 
is completely missing in our results. Moreover the calculated magnitude of
this term above 40~\AA\ is remarkably underestimated. 
The fact suggests that folding effects give only a minor contribution,
while the origin of the behavior of $\Delta n_{bg}$
must be attributed to local field effects, 
as already suggested in Ref.\onlinecite{sirenko99}.
Similar considerations can be made for the GaAs/vacuum system.
 
The magnitude of the background birefringence related to local field
effects can be easily estimated in the case of long wavelength of incident
radiation and not too small superlattice periods $n$. If boundaries are assumed
abrupt and the constituent materials are supposed to conserve their bulk
dielectric functions up to the interfaces, we can apply approximate
expressions for $\epsilon_{\perp}$ and $\epsilon_{\parallel}$ 
in terms of bulk constituent scalar dielectric functions, 
$\epsilon_1$ and $\epsilon_2$ \cite{agranovich85}:
\be
\epsilon_{\perp} \left( \omega \right) = \frac{1}{l_1+l_2} 
\left( \epsilon_1 \left( \omega \right) l_1 + \epsilon_2 
\left( \omega \right) l_2 \right) \,,
\label{eperp}
\ee
\be
\epsilon^{-1}_{\parallel} \left( \omega \right) =  
\frac{1}{l_1+l_2} 
\left( \frac{l_1}{\epsilon_1 \left( \omega \right)} + 
\frac{l_2}{\epsilon_2 
\left( \omega \right)}  \right) \,,
\label{epar}
\ee
where $l_1$, $l_2$ are the layer thicknesses of the two different materials. 
Using our calculated values of bulk static dielectric constants in
expressions (\ref{eperp}) and (\ref{epar}) we find a rough 
estimate $\Delta n\simeq 0.03$ for the local field contribution to
birefringence. This value is much larger than the calculated values 
in Fig.~\ref{fitpara}, for all superlattice periods $n$, and it is of the same order
of magnitude as the the experimental results at intermediate $n$.
Only for small periods $n$, when relations (\ref{eperp}) and (\ref{epar})
do not hold, the  local field correction should become small.
Precise calculations of the optical properties including
local-field effects are obviously required to clarify this point.

\section{Summary and discussion}

The energy bands and optical constants of (001)-oriented
(GaAs)$_n$/(AlAs)$_n$ and (GaAs)$_n$/(vacuum)$_n$ superlattices 
with $n$ from 4 to 20 have been calculated by the LCBB
method introduced in Ref.~\onlinecite{wang97}. This approach,
in which the electronic states of the superlattice are expanded
in the basis of bulk states calculated by empirical pseudopotentials,
is found to be adequate and practical for superlattices with small to
intermediate period; it is particularly useful for calculating
how the optical spectra of the bulk materials are modified
upon confinement.

Quantum-confined induced shifts of the critical point energies 
are calculated for both kinds of superlattices and are found 
to be larger for the GaAs/vacuum system, where coupling between
different GaAs layers is only due to quantum-mechanical tunneling
and has a negligible effect.
For both GaAs/AlAs and GaAs/vacuum superlattices, the $E_1$ peak 
in the absorption spectrum splits into two peaks with different blue
shifts for decreasing superlattice period.
This result agrees with the observations of Ref.~\onlinecite{garriga87}
on GaAs/AlAs superlattices, and is attributed to a symmetry splitting
of the valence bands at the $L$ point. The blue shifts are again larger
for the GaAs/vacuum system. The $E_2$ transition instead is found to be
split for large-period GaAs/AlAs superlattices, where the electronic
states of the bulk are confined in each layer and the absorption spectrum
is the superposition of the two bulk ones; the energy of the $E_2$ peak
depends weakly on the superlattice period.
The average or Penn gap, defined as the first zero of the real part
of the dielectric constant, does not depend on superlattice period,
confirming the expectation that a blue shift at the lower absorption edges
is compensated by red shifts in the upper parts of the absorption
spectrum. The band contribution to linear birefringence of GaAs/AlAs
superlattices is calculated and compared with recent experimental results 
of Ref.~\onlinecite{sirenko99}.
The zero-frequency birefringence is found to be much smaller than
the experimental results: the observed static birefringence
is attributed to local-field effects as already suggested.\cite{sirenko99}
The frequency-dependent part of the birefringence arising from
band folding and quantum confinement increases with decreasing
superlattice period as found in the experiment, although the calculated
values are smaller.

The present work can be extended in several directions.
Within the LCBB scheme relying on empirical pseudopotentials,
a more precise calculation of the optical properties requires
the inclusion of spin-orbit interaction and of local-field effects:
the latter are particularly important to account for
birefringence of the superlattice. 
A complete description of the $E_1$ peak requires the inclusion
of excitonic effects, which account for half of the oscillator
strength of the transition in the bulk.
For superlattices with very short period, the expansion over bulk states
becomes unreliable (unless using a very large basis)
and an ab-initio approach is required:
it would be interesting to compare the results of ab-initio
and LCBB calculations for superlattices with a period $n$
from 4 to 8, say, where ab-initio methods are feasible
and the LCBB approach is still meaningful.

\section*{Acknowledgements}

The authors are grateful to M.\ Cardona, R.\ Del Sole,
A.\ Stella and P.\ Tognini for several helpful suggestions and comments
and to L.\ Reining and N.\ Vast for a critical reading of the manuscript.

%%%%%%%%%%%%%%%%%%%%%%%%%%%%%%%%%%%%%%%%%%%%%%%%%%%%%%%%%%%%%%%%%%%%%%%%%%%%%%

%References

%\end{multicols}
\bfig[h]
\begin{center}
 \epsfig{file=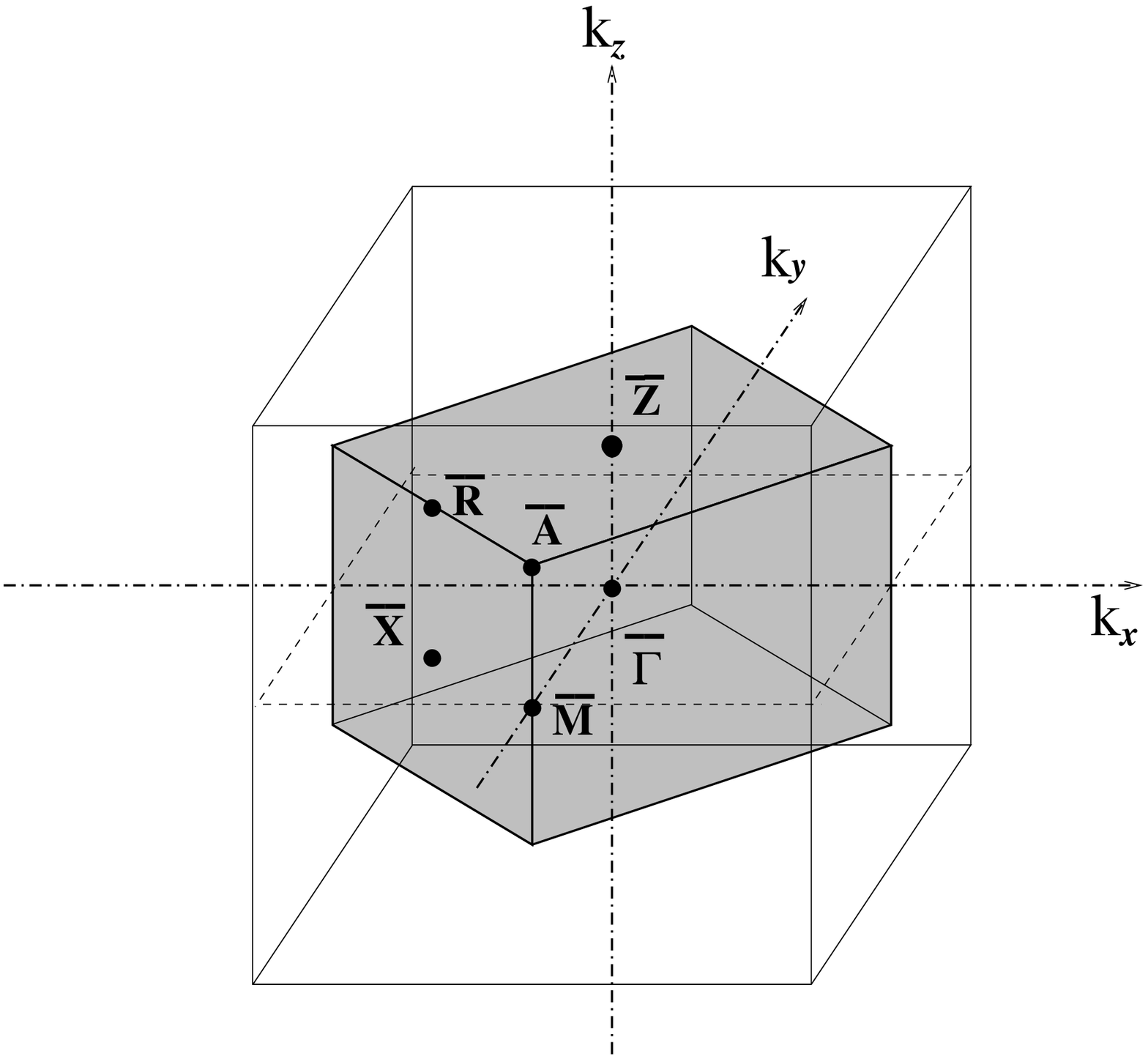, width=6truecm}
 \end{center} 
 \caption{\label{tetrbz} Brillouin zone for simple tetragonal 
(GaAs)$_n/$(AlAs)$_n$ and (GaAs)$_n/$(vacuum)$_n$ (001) 
superlattices,  included in bulk   conventional cubic cell. 
The figure shows high symmetry points positions.} 
\efig

\bfig[h]
\begin{center}
 \epsfig{file=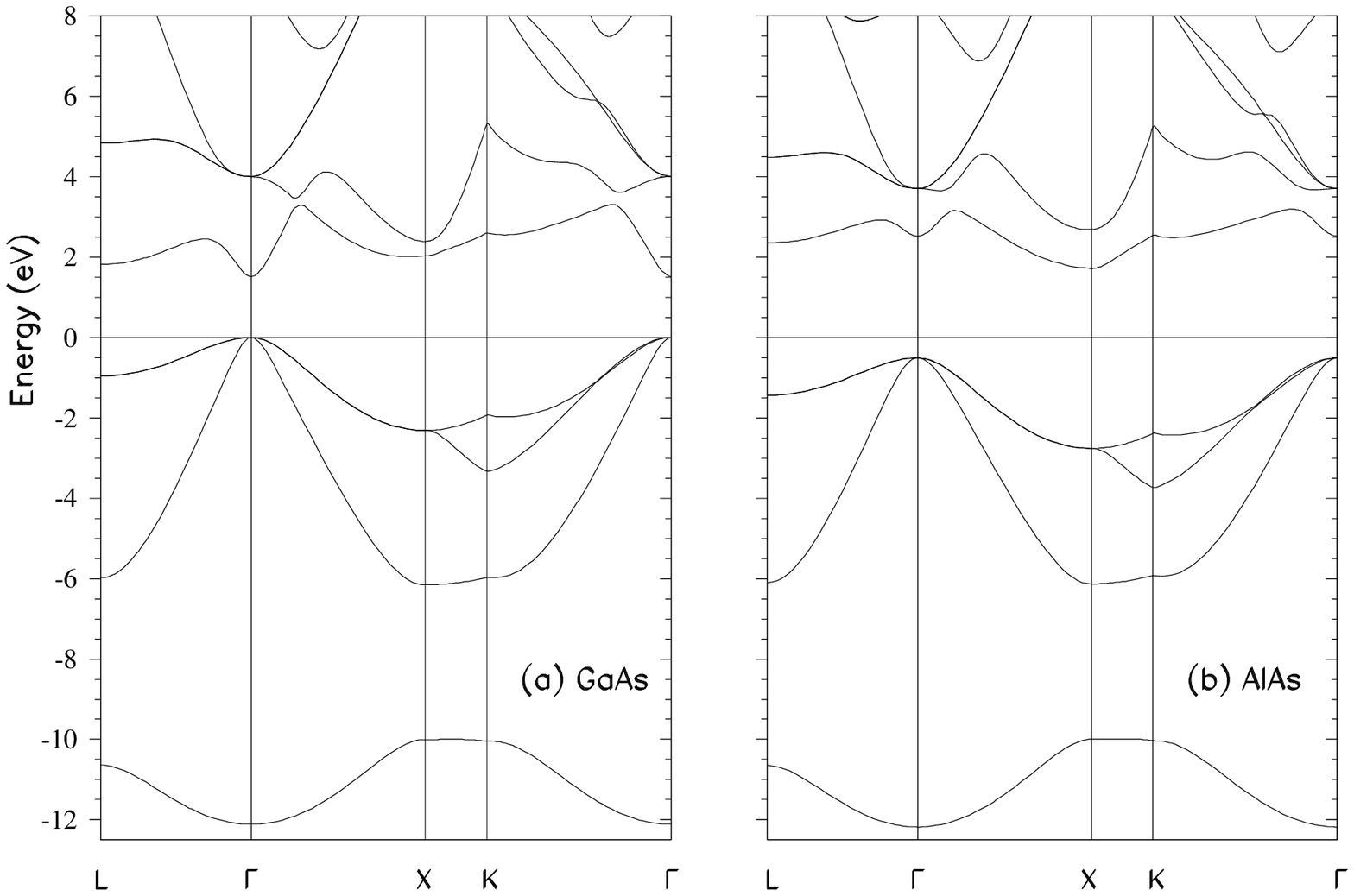, width=8.5truecm}
 \end{center} 
 \caption{\label{bulkbands} Bulk band structure of (a) GaAs and (b) AlAs
 crystals along the high symmetry directions,
 obtained by empirical pseudopotentials
 in Ref.~\protect\onlinecite{mader}.
 The energy zero is always taken at 
 the valence-band maximum of bulk GaAs and the valence band offset 
 between GaAs and AlAs is 0.5 eV.} 
\efig

\bfig[h]
\begin{center}
 \epsfig{file=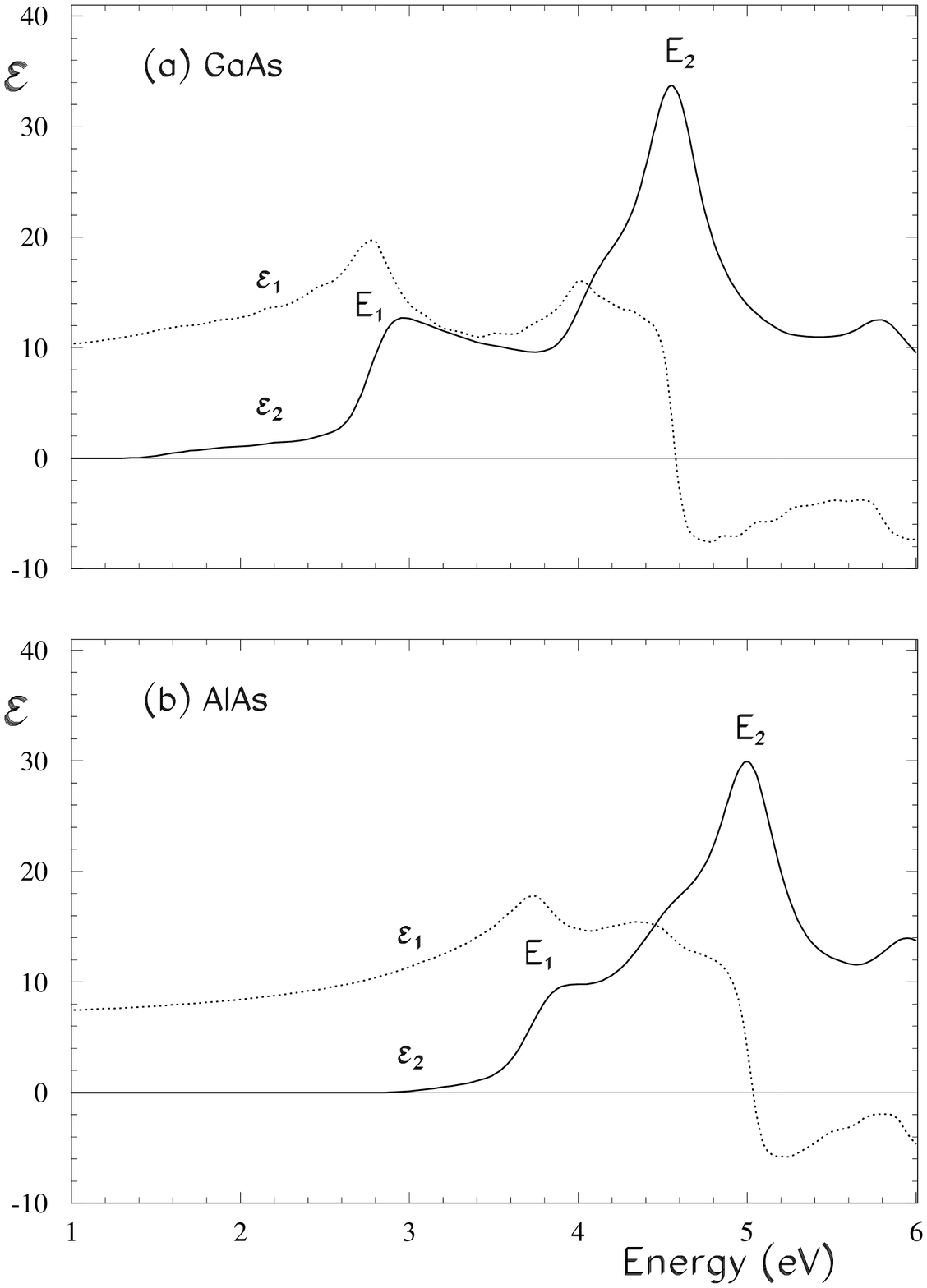, width=8.5truecm}
 \end{center} 
 \caption{\label{e2bulk} Calculated dielectric function (real and imaginary
 parts) of (a) GaAs and (b) AlAs.} 
\efig

 \bfig[h]
\begin{center}
 \epsfig{file=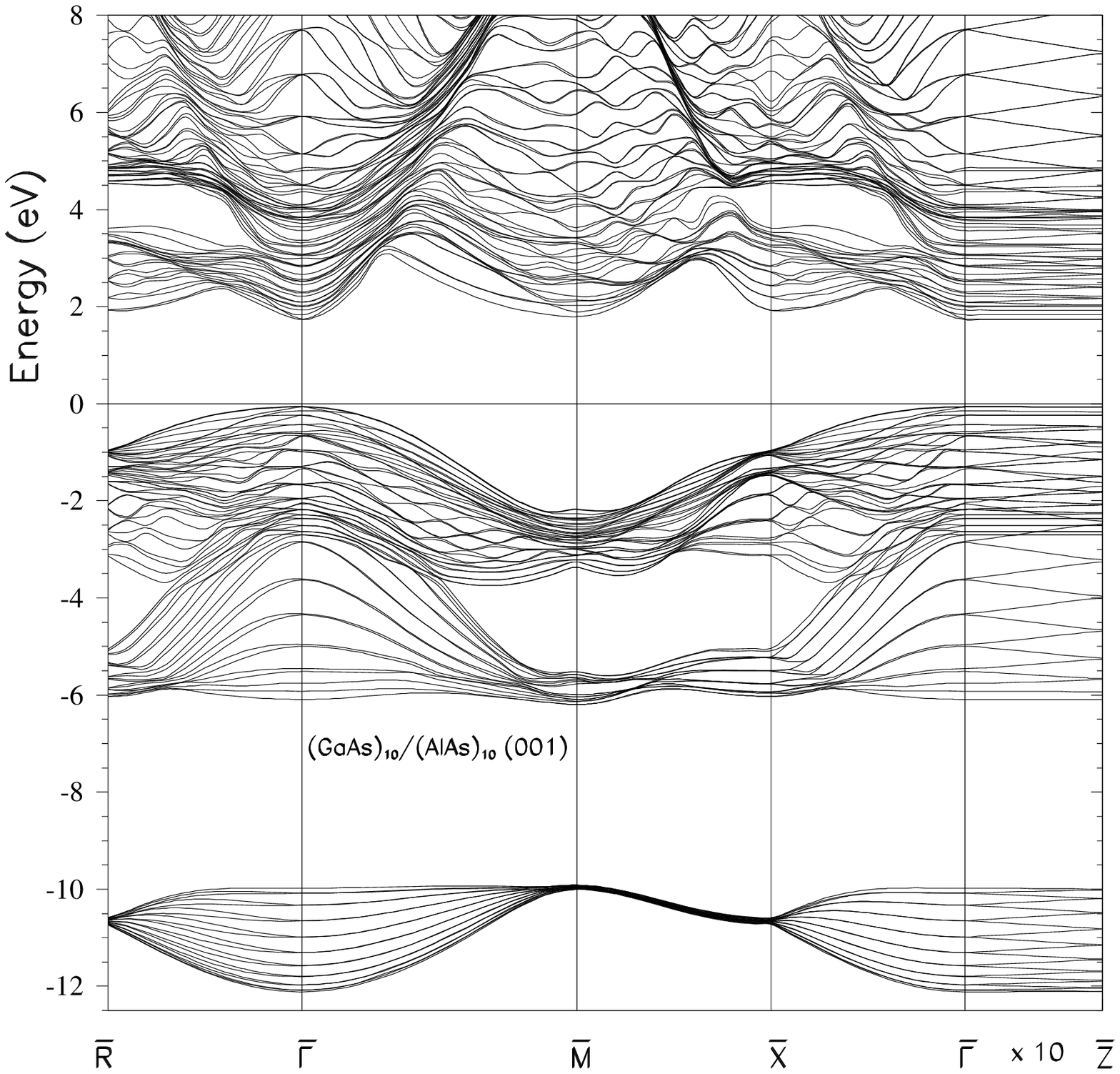, width=8.5truecm}
 \end{center} 
 \caption{\label{SLbandsgaal} Band structure of a (GaAs)$_{10}/$(AlAs)$_{10}$ 
 (001) superlattice along the high symmetry directions.
 The length of the $\bar \Gamma - \bar Z$ line is multiplied by ten for clarity.
  The energy zero is taken at the bulk
 GaAs valence-band maximum.} 
\efig

\bfig[h]
\begin{center}
 \epsfig{file=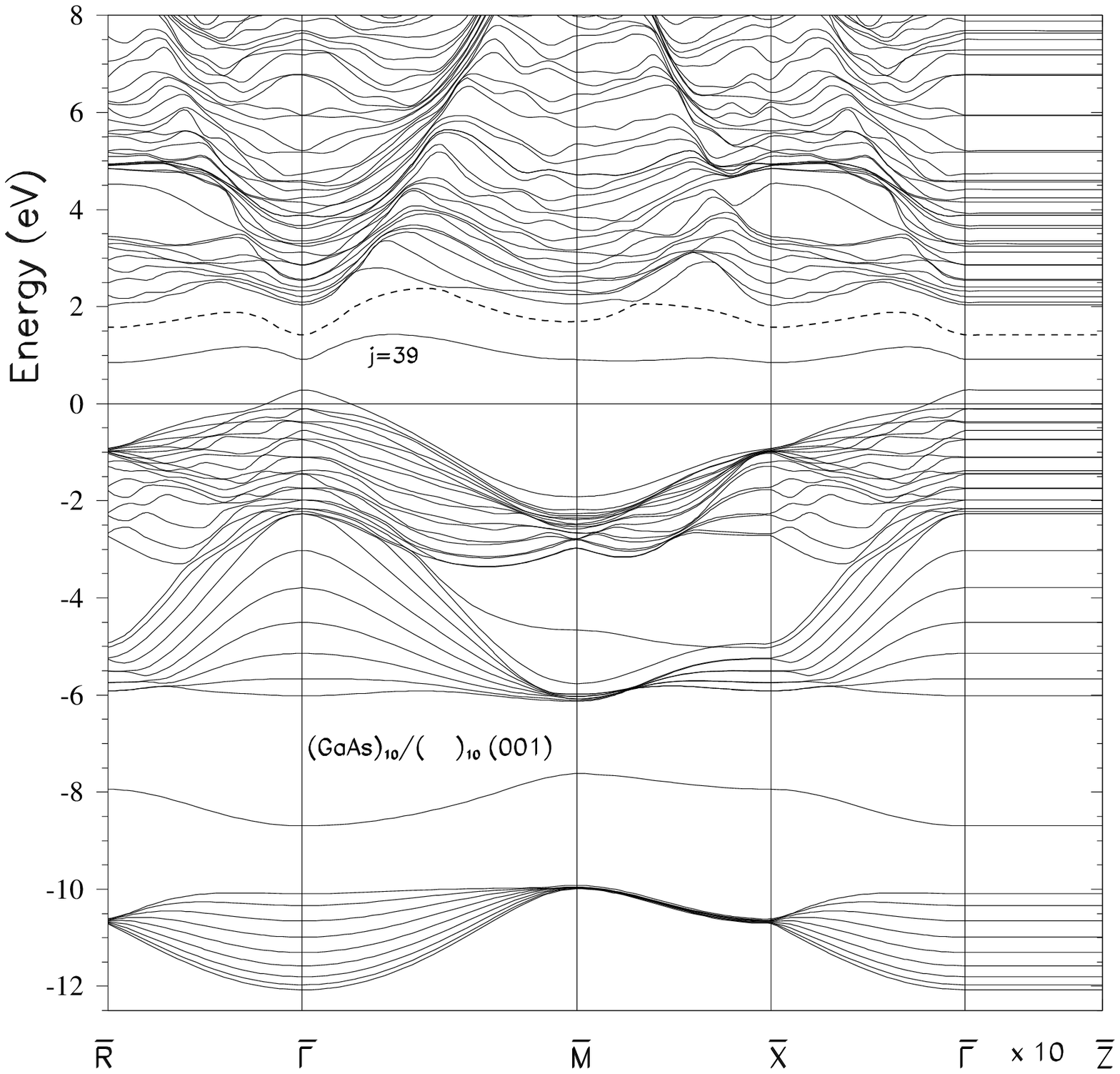, width=8.5truecm}
 \end{center} 
 \caption{\label{SLbandsvacu} Band structure of a (GaAs)$_{10}/$(vacuum)$_{10}$ 
 (001) superlattice along the high symmetry directions.
The energy zero is taken at the bulk
 GaAs valence-band maximum. 
 The uppermost occupied band is number 40 (dotted).} 
\efig

\bfig[h]
\begin{center}
 \epsfig{file=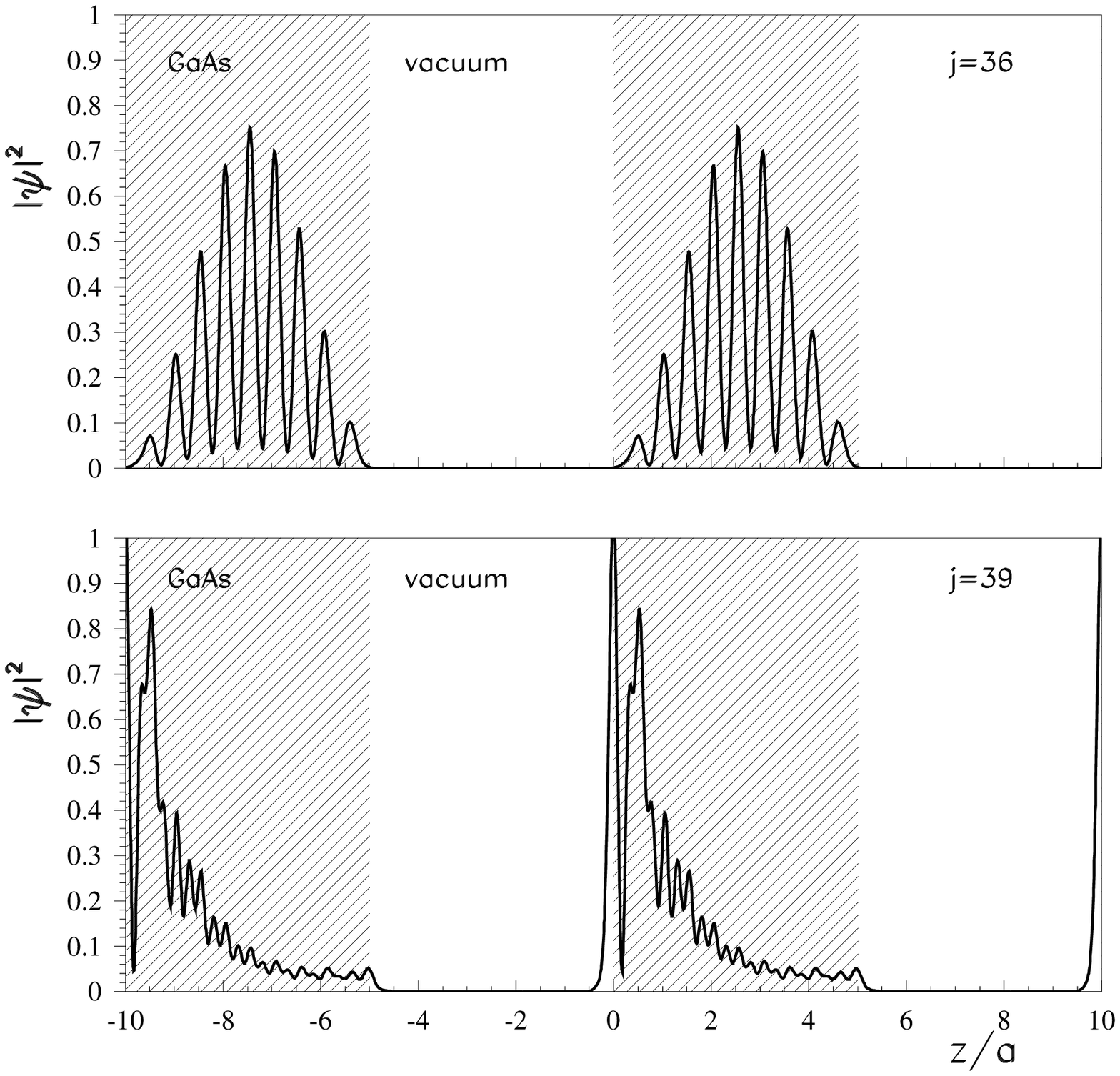, width=8.5truecm}
 \end{center} 
 \caption{\label{surface} Planar averaged probability along the growth
 direction $z$ to find an electron at $\Gamma$ 
 in the 36th and the 39th band of a (GaAs)$_{10}/$(vacuum)$_{10}$ (001) 
superlattice.} 
\efig

\bfig[h]
\begin{center}
 \epsfig{file=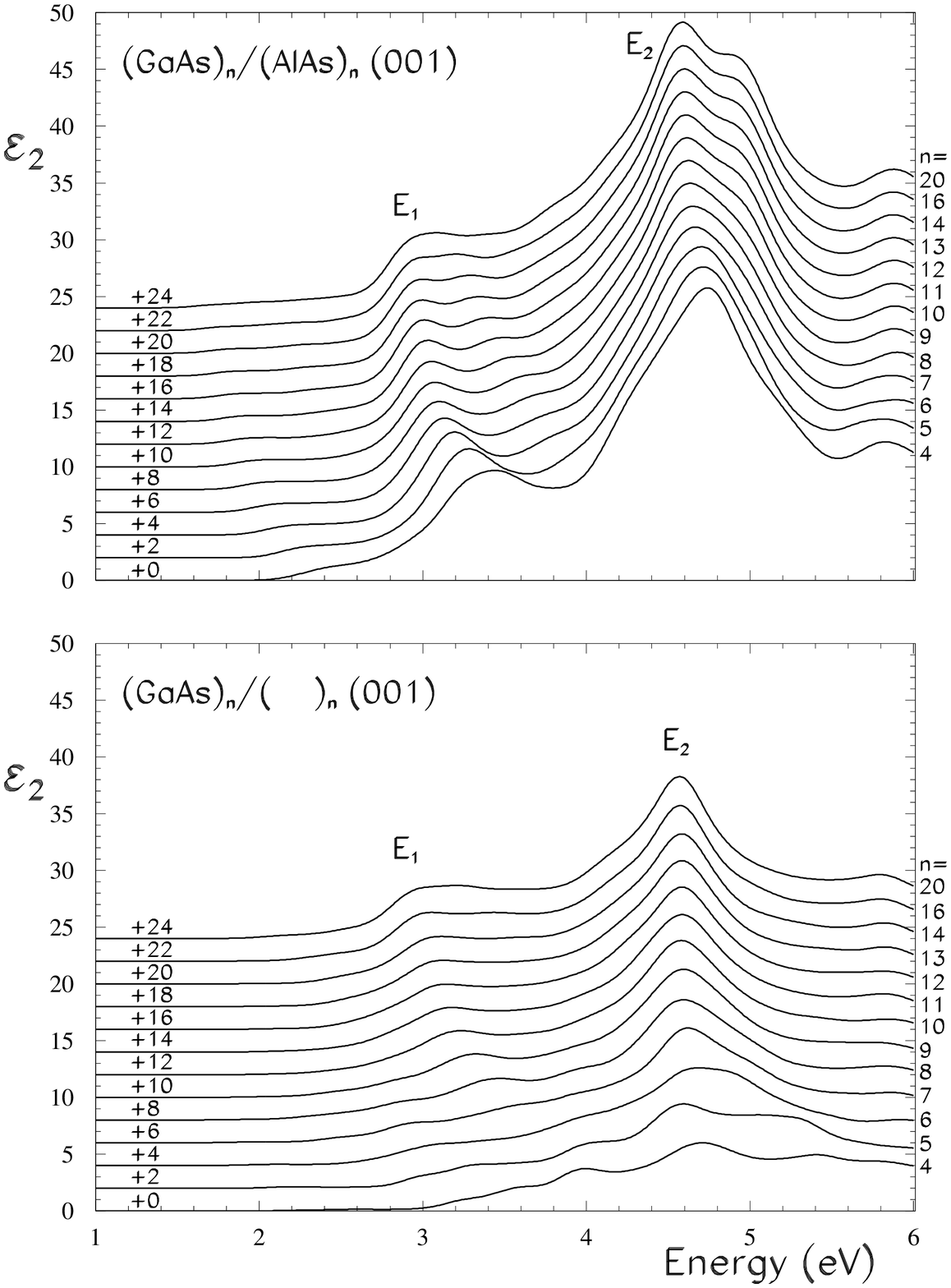, width=8truecm}
 \end{center} 
 \caption{\label{eps2SL} Imaginary part of the dielectric function
 for (GaAs)$_{n}/$(AlAs)$_{n}$ and (GaAs)$_{n}/$(vacuum)$_{n}$ superlattices,
 for different values of the period $n$. 
 Different curves are offset for clarity.
 $E_1$ splitting cannnot be easily seen in the figure: 
 the peak positions have been
 determined by an enlargement of the spectral region of interest.} 
\efig

\bfig[h]
\begin{center}
 \epsfig{file=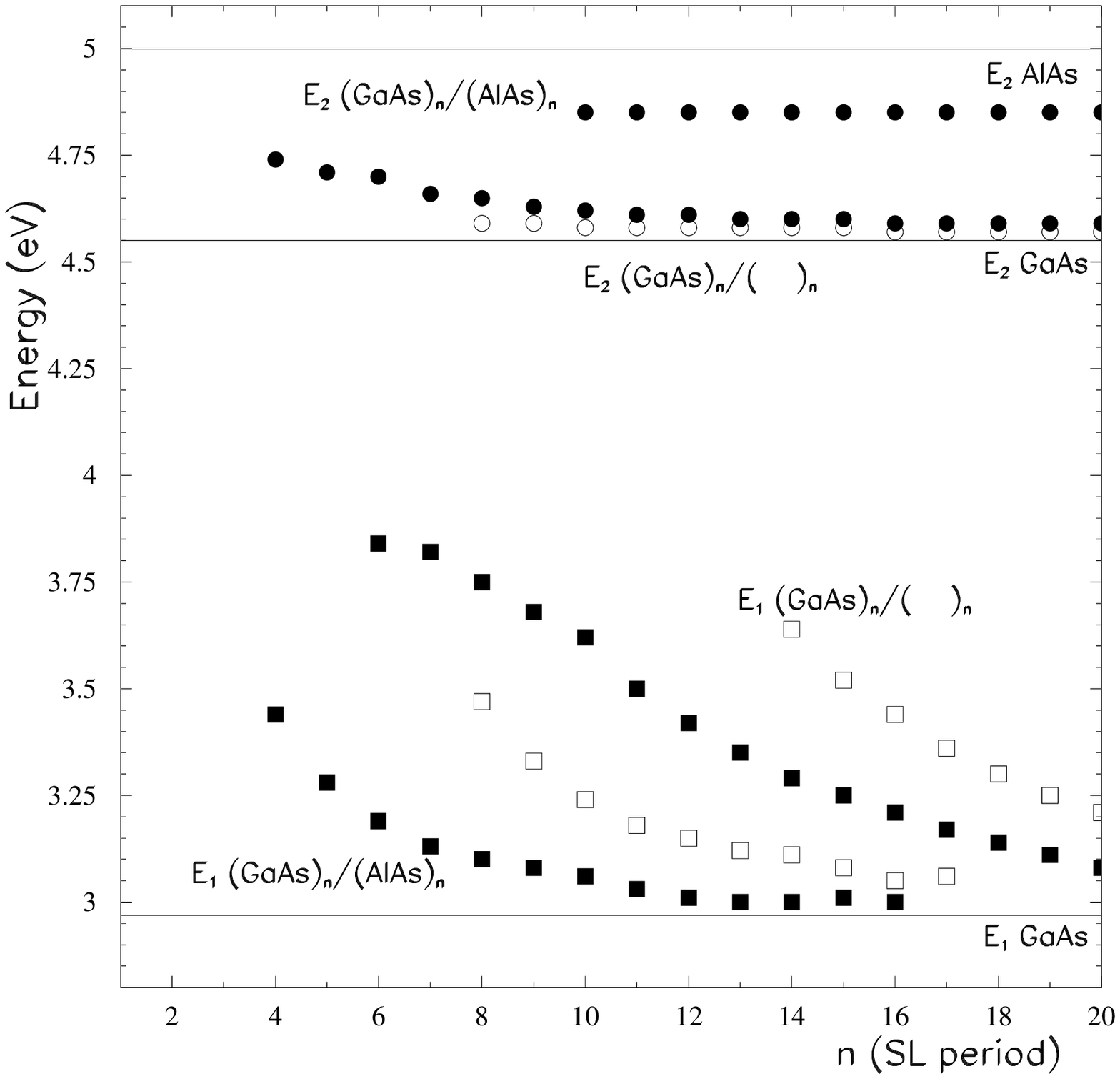, width=8truecm}
 \end{center} 
 \caption{\label{picchi} $E_1$ and $E_2$ peak positions for
 GaAs/AlAs (closed symbols) and GaAs/vacuum (open symbols)
 as a function of superlattice period $n$. The horizontal lines represent 
the peak energies in the bulk.} 
\efig

\bfig[h]
\begin{center}
 \epsfig{file=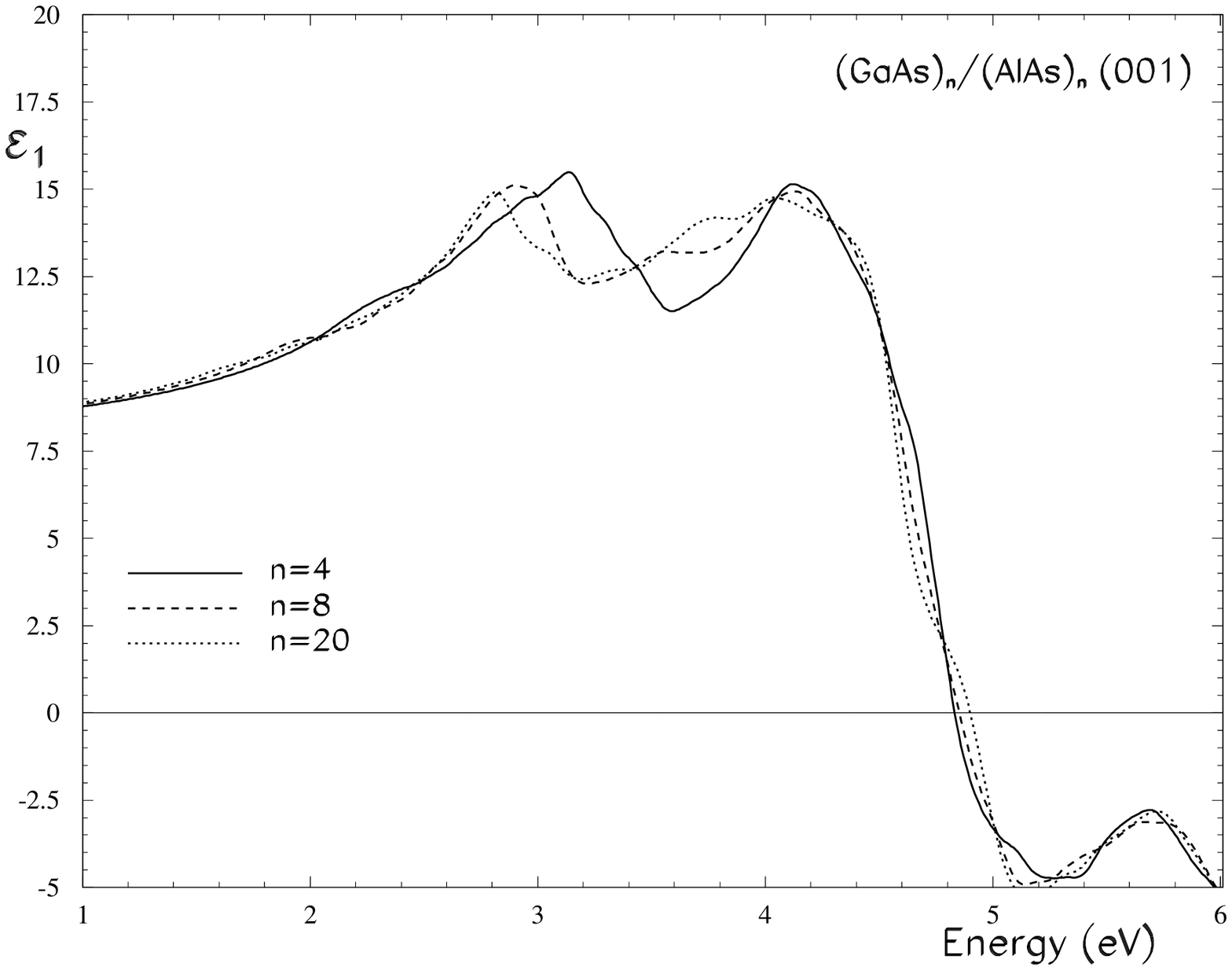, width=8.5truecm}
 \end{center} 
 \caption{\label{e1} Real part of dielectric function
 for (GaAs)$_{n}/$(AlAs)$_{n}$ (001) superlattices, for
 different values of the superlattice period $n$.} 
\efig

\bfig[h] 
\begin{center}
 \epsfig{file=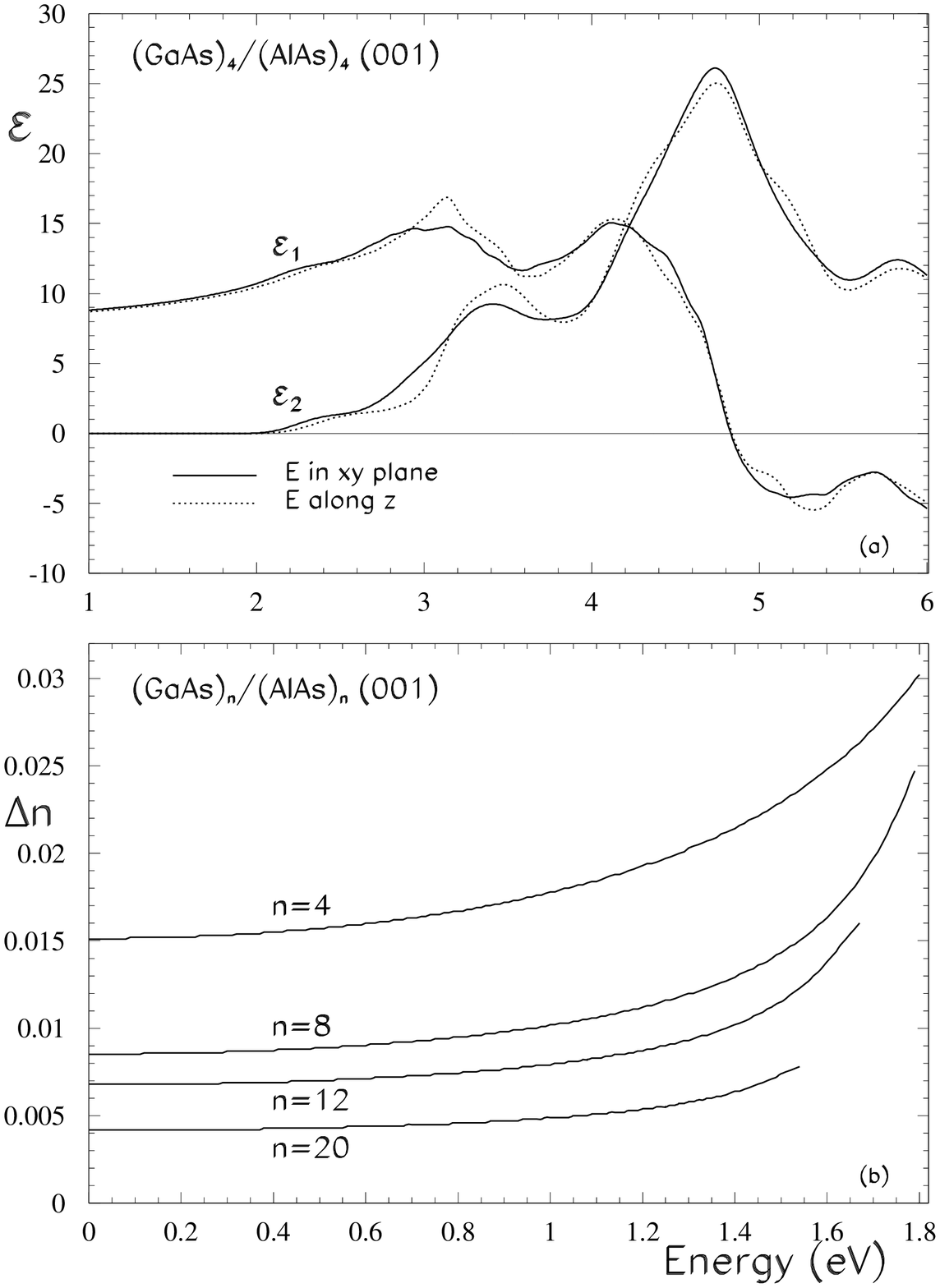, width=8.5truecm}
 \end{center} 
 \caption{\label{anis} (a) Components of the dielectric tensor 
 (real and imaginary parts) for a (GaAs)$_{4}/$(AlAs)$_{4}$ 
 (001) superlattice
 and (b) linear birefringence
 $\Delta n=(\epsilon_{\perp})^{1/2}-(\epsilon_{\parallel})^{1/2}$
 for (GaAs)$_n$/(AlAs)$_n$ (001) superlattices.
} 
\efig

\bfig[h] 
\begin{center}
 \epsfig{file=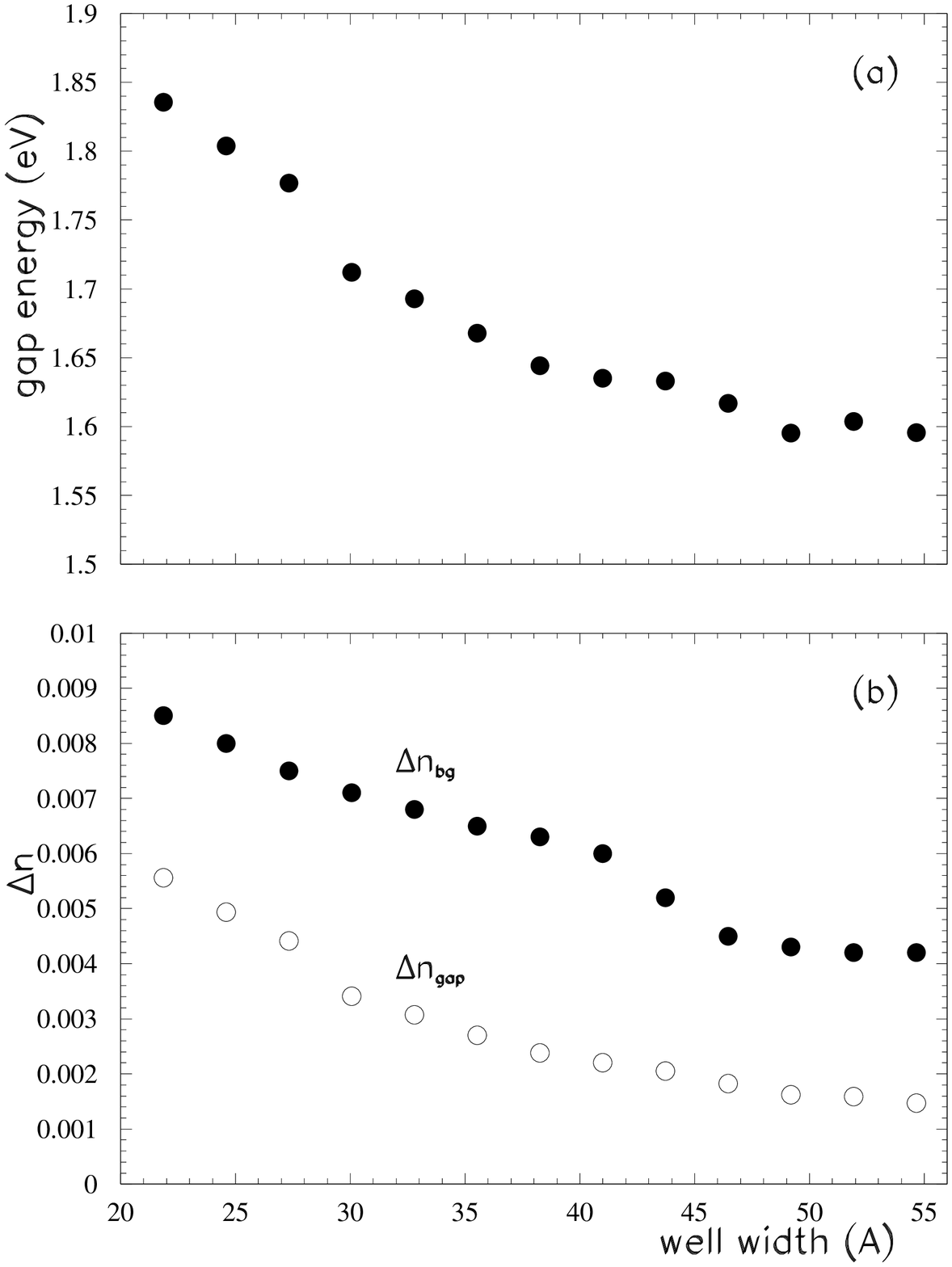, width=8.5truecm}
 \end{center} 
 \caption{\label{fitpara} Parameters of equation (\ref{fit}) as a function
 of well width (half of a superlattice period): 
(a) energy gap, (b) $\Delta n_{gap}$ and $\Delta n_{bg}$. 
 The results can be compared with the analogous experimental graphs in
 Ref.~\protect\onlinecite{sirenko99}. }
\efig
\mybeginwide

\begin{table}
\caption{ \label{tabella} 
Comparison of energy levels (in eV) at the symmetry points for
(GaAs)$_n$/(AlAs)$_n$ superlattices with period $n=6$ and $n=10$.
We show highest valence levels and lowest conduction levels: the energy zero
is taken at the valence band maximum.
Four different choices  are considered for the basis set:
1) five  GaAs bands for $n_b$ from 1 to 5, no AlAs bands;
2) eight GaAs bands for $n_b$ from 1 to 8, no AlAs bands;
3) eight GaAs bands for $n_b$ from 1 to 8, the 5th band of AlAs;
4) eight GaAs bands for $n_b$ from 1 to 8, 
   four  AlAs bands for $n_b$ from 5 to 8.
}
\begin{center}
\begin{tabular}{c|cccccccccc}
 basis & \multicolumn{5}{c}{valence band maximum} & 
 \multicolumn{5}{c}{conduction band minimum} \\  
       &  $\bar \Gamma$ &  $ \bar Z$ & $\bar R$ & $\bar M$  &$\bar X$  & 
        $\bar \Gamma$ &  $ \bar Z$ & $\bar R$ & $\bar M$  &$\bar X$ \\
 \hline
 n=10 & & & & & & & & & & \\
 1) &  0.0000 & -0.0004 & -0.8992  & -2.1127  & -0.8992   
 & 1.8318 & 1.8303   & 2.0069  &  1.9048 &  2.0063 \\
 2) &  0.0000 &  -0.0004  &   -0.8990  &   -2.1129   &   -0.8990 & 
  1.8173  &   1.8190  &   2.0057   &  1.9014    &  2.0049 \\
 3) &   0.0000  &  -0.0004 &   -0.8990 &   -2.1131  &   -0.8990  &
   1.7949  &    1.8008   &   1.9911  &    1.8588  &     1.9881 \\
 4) &  0.0000  &  -0.0003  &  -0.8985  &  -2.1130  &   -0.8985 &
     1.7884  &   1.7957  &   1.9904   & 1.8590  &    1.9874 \\
 \hline  
 n=6 & & & & & & & & & & \\
 1) &   0.0000  &   -0.0118  &  -0.8488  &  -2.0968   &  -0.8489  &
 1.9766  &    1.9940   &  2.1730  &   1.9961   &   2.1561 \\
 2)   &  0.0000   &  -0.0115  &  -0.8481  &  -2.0970  &   -0.8482  &
 1.9526   &   1.9567  &   2.1694   &  1.9940   &   2.1535 \\ 
 3) &   0.0000   &   -0.0115  &   -0.8481  &   -2.0973   &   -0.8482 & 
  1.9062  &     1.9109   &   2.1451  &    1.9618   &    2.1053 \\
 4)  &   0.0000  &    -0.0113 &   -0.8468 &   -2.0967  &   -0.8468   & 
  1.9042  &    1.9098  &   2.1440   &  1.9626   &   2.1043 \\
\end{tabular}
\end{center}

\end{table}


\begin{references}
\bibitem{note_period} The period of the superlattice is related to the
number $n$ of monolayers of each material by $d=2n\delta$,
where $\delta$ is the distance between neighboring planes of the same atom.
For (001)-grown superlattices, as considered here, $d=na$.
\bibitem{e0prime} What we call $E_2$ peak may in fact contain
weak contributions from the $E_0'$ transition, which can
be clearly resolved only in derivative spectra.
See e.g.\ Refs.~\protect\onlinecite{chelikovsky-cohen,garriga87}.
\end{references}
\end{document}